\documentclass[aip,cha,reprint,nofootinbib,a4paper]{revtex4-1}
\usepackage[compact]{titlesec}
\titleformat*{\section}{\large\bfseries}
\titlespacing{\section}{0pt}{8pt}{0pt}

\usepackage[utf8]{inputenc}
\usepackage{amsmath, amsthm, amssymb}
\usepackage{enumitem}
\usepackage{graphicx}
\usepackage{times}
\usepackage{dcolumn} 
\usepackage{url}
\usepackage{xcolor}
\usepackage{soul}
\DeclareMathAlphabet{\altmathcal}{OMS}{cmsy}{m}{n}

\usepackage[labelsep=space]{caption}

\DeclareCaptionFont{helv}{\fontfamily{helvet}\selectfont}

\begin{document}


\title{Beyond network centrality: Individual-level behavioral traits for predicting information superspreaders in social media}

\author{Fang Zhou$^{1,3}$}

\author{Linyuan L\"u$^{2,1,*}$}
\author{Jianguo Liu$^{4}$}
\author{Manuel Sebastian Mariani$^{1,5,*}$}

\footnotetext[1]{Institute of Fundamental and Frontier Sciences, University of Electronic Science and Technology of China, Chengdu 610054, P. R. China}
\footnotetext[2]{School of Cyber Science and Technology, University of Science and Technology of China, Hefei, 230026, P. R. China}
\footnotetext[3]{Yangtze Delta Region Institute (Huzhou), University of Electronic Science and Technology of China, Huzhou 313001, P. R. China}
\footnotetext[4]{Institute of Accounting and Finance, Shanghai University of Finance and Economics, 200433, P. R. China}
\footnotetext[5]{URPP Social Networks, Universit\"at Z\"urich, Switzerland}
\renewcommand{\thefootnote}{*}
\footnotetext{e-mail:linyuan.lv@uestc.edu.cn; manuel.mariani@business.uzh.ch}

\begin{abstract}
Understanding the heterogeneous role of individuals in large-scale information spreading is essential to manage online behavior as well as its potential offline consequences. To this end, most existing studies from diverse research domains focus on the disproportionate role played by highly-connected ``hub" individuals. However, we demonstrate here that information superspreaders in online social media are best understood and predicted by simultaneously considering two individual-level behavioral traits: influence and susceptibility. Specifically, we derive a nonlinear network-based algorithm to quantify individuals' influence and susceptibility from multiple spreading event data. By applying the algorithm to large-scale data from Twitter and Weibo, we demonstrate that individuals' estimated influence and susceptibility scores enable predictions of future superspreaders above and beyond network centrality, and reveal new insights on the network position of the superspreaders.     
\end{abstract}

\maketitle

\section*{INTRODUCTION}
\begin{figure*}[t]
    \centering
    \includegraphics[scale=0.36]{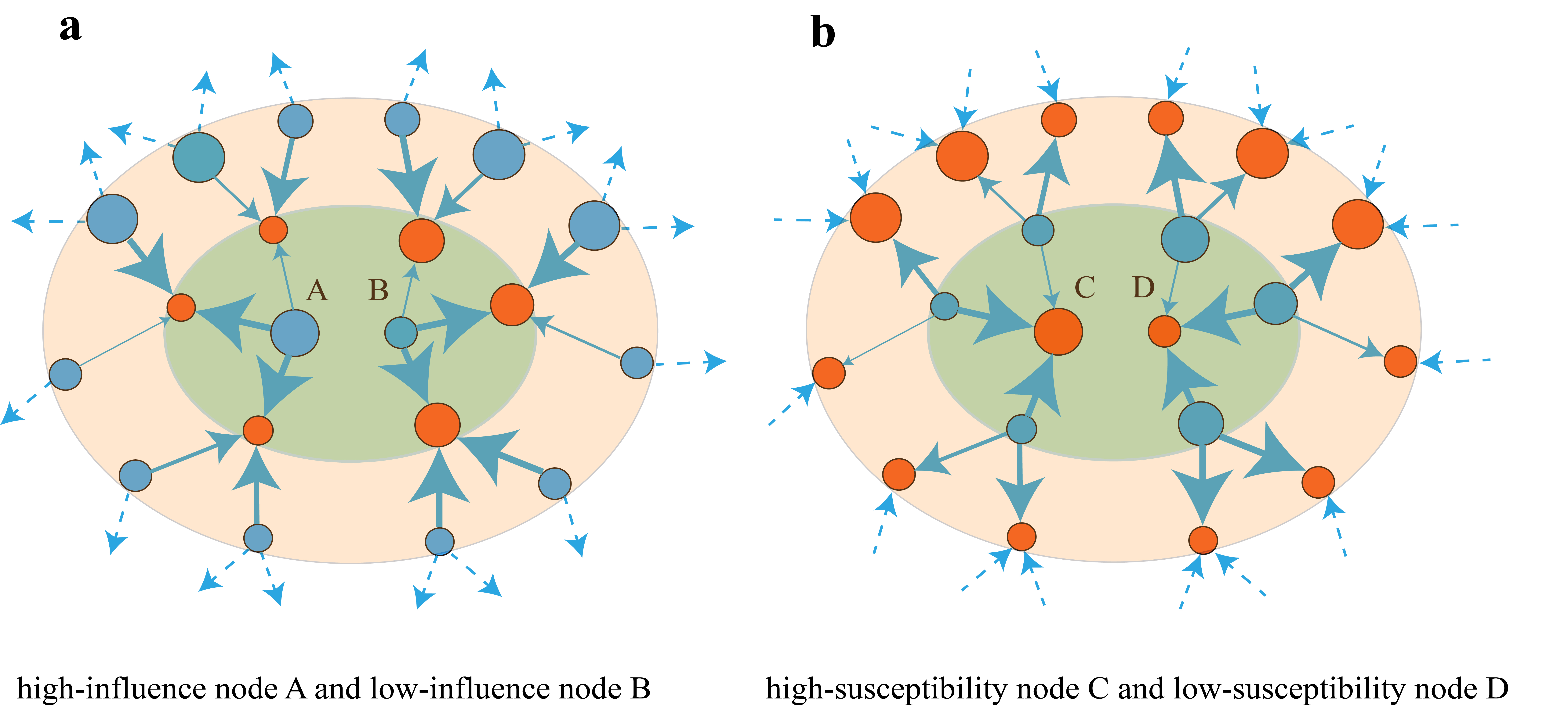}
    \caption{{\bf Quantifying individual influence and susceptibility from observational data.} To illustrate the intuition behind the proposed influence-susceptibility algorithm, in panel {\bf a}, we show how two nodes (A, B) with the same number of outgoing links and the same number of propagation events can achieve a widely different influence score $I$. The size of orange (blue) nodes denotes their susceptibility score $S$ (influence score $I$); the thickness of the arrows represents the number of propagation events. Both A and B have three outgoing links, meaning that they influence three nodes. However, A's neighbors have a lower susceptibility than node B's neighbors, because they are influenced fewer times by their other neighbors (the blue nodes in the outer-most shell). As A influences less susceptible individuals than B does, according to the IS algorithm, A is more influential than B. With a similar argument, in panel {\bf b}, node C is more susceptible than D.
   }
    \label{fig:illustration}
\end{figure*}

Modern communication technologies and online social media have radically transformed the way information propagates in human societies. 
Online information simultaneously influences and is influenced by offline behavior at an unprecedented pace, which offers both potential benefits and threats to mankind\cite{bak2021stewardship}.
Potential benefits include accelerating human cooperation during emergencies\cite{lacassin2020rapid}, promoting community-minded\cite{chen2010social} and healthy behavior\cite{centola2018behavior}, and increasing public trust in science\cite{huber2019fostering}.
On the other hand, uncontrolled online information flows can accelerate the spreading of dangerous misinformation\cite{lazer2018science,shao2018spread,bovet2019influence,guess2020exposure,gallotti2020assessing}, exacerbate polarization in society\cite{stella2018bots,johnson2020online,baumann2020modeling,medo2021fragility}, and even trigger civil unrest\cite{gonzalez2011dynamics,mooijman2018moralization}.

Given the significant implications of online information spreading processes for global collective behavior, a major cross-disciplinary challenge is to provide mechanistic insights into the structure and dynamics of the spreading processes\cite{aral2018social,hu2018local,wang2019anomalous,shi2019totally,tang2020predictability,xie2021detecting,zhou2020realistic,juul2021comparing}, which could be beneficial for predicting and managing their large-scale consequences\cite{bak2021stewardship}. 
In this respect, detecting superspreaders -- individuals capable of triggering large-scale information cascades in social media -- is crucial to diverse decision-makers, ranging from firms aiming to promote awareness for a new product\cite{hinz2011seeding,muller2019effect} to policymakers and regulators seeking to design effective interventions to curb the large-scale diffusion of misinformation\cite{budak2011limiting,shao2018spread,bovet2019influence,grinberg2019fake}.
To shed light on the micro-level mechanisms behind information propagation in social media, we ask: To what extent can we predict future information superspreaders? Is an individual's future spreading ability better predicted by her position in the social network or by her behavioral traits? 

No consensus has been reached regarding the answers to these questions. Building on seminal opinion leadership and diffusion theories\cite{katz1955personal,rogers2010diffusion}, researchers as diverse as sociologists, economists, physicists, and computer scientists have tackled the problem by linking the micro-level properties of individuals and social ties with the large-scale spreading of information and behavior. Among them, many studies supported the hypothesis that a minority of influential individuals exert a disproportionately significant impact on online and offline information spreading, and that these ``influentials" can be detected via their centrality in the social networks they are embedded in. These works relied on diverse methods, ranging from observational data analysis\cite{goldenberg2009role,muller2019effect}, computational diffusion models\cite{domingos2001mining,kempe2003maximizing,kitsak2010identification,pei2014searching,lu2016vital,hu2018local,zhou2019fast} to field experiments\cite{hinz2011seeding,banerjee2013diffusion}.

Yet some works disagreed with the above conclusion\cite{watts2007influentials,galeotti2009influencing,mariani2020wisdom,rossman2021network}, holding that centrality metrics alone do not reveal how information and behavior propagate from a given individual to her peers. Focusing on the diffusion of a single movie product, a seminal field experiment\cite{aral2012identifying} found that peer-to-peer contagion in social media is jointly affected by two individual-level behavioral properties: \textit{influence}, namely how strongly people influence their neighbors; and \textit{susceptibility}, how easily people are influenced by their neighbors\cite{aral2012identifying}.  They found that incorporating influence and susceptibility into agent-based diffusion models generates radically different predictions of information spreading processes than traditional diffusion models and centrality metrics\cite{aral2018social}.
However, when comparing these findings with those from the literature on social hubs, it remains unclear whether, compared to network-based characteristics such as centrality, individual-level behavioral traits are more accurate predictors of information superspreaders in empirical diffusion processes.


Here, we demonstrate that information superspreaders in social media are better predicted by individuals' behavioral traits than by individuals' centrality in the diffusion network. To this end, we first develop an algorithm that simultaneously quantifies individuals’ influence and susceptibility to influence from multiple spreading events. Based on an empirical influence model\cite{aral2018social}, we find that individual influence and susceptibility follow coupled non-linear, network-dependent equations.  We also validate the algorithm's ability to reconstruct individual influence and susceptibility in synthetic spreading data, even when the input data are incomplete. We further show that the algorithm's scores in empirical data respect expected properties of the network position of highly-influential and susceptible individuals\cite{aral2012identifying}. 

By applying the new algorithm to social media data collected from Twitter and Weibo, we reveal previously unknown assortativity properties of individual centrality, influence, and susceptibility. Differently from recent results\cite{zhou2020realistic}, our findings indicate that individuals' influence and susceptibility are not necessarily correlated with their centrality.
Crucially, we show that compared to network structural metrics (such as centrality\cite{zhou2020realistic}) and users' past success\cite{bakshy2011everyone}, knowledge of both individuals' influence and susceptibility can substantially improve the out-of-sample detection of superspreaders. Feature importance analysis reveals that beyond users' past success, the most important predictors of being a future superspreader include the users' influence as well as the influence and susceptibility of the influenced users. In particular, linear regression analysis suggests that the superspreaders are not disproportionately central in the social network, but they exhibit more high-contagion links (i.e., directed links that connect highly influential with highly susceptible users) and they tend to influence more influential users.


Compared to the many studies on the role of social networks in spreading processes, this collection of findings undermines the common assumption that network structural metrics are necessary to reliably predict future superspreaders. Our findings indicate that more accurate predictions are obtained by integrating individual-level behavioral traits and network properties. Future studies may seek to extend this conclusion from the online information spreading problem studied here to the complex diffusion of offline behaviors\cite{centola2018behavior}, which is highly relevant to societal challenges such as the promotion of sustainable consumption and vaccination.

\section*{Results}

\textbf{Algorithm derivation and validation.}
We consider a set of $N$ individuals that form the nodes of a directed diffusion network represented by its adjacency matrix, $\mathbf{A}$.
We consider a recent diffusion model~\cite{aral2018social} -- which we refer to as the \textit{empirical diffusion model} -- that assumes that the peer-to-peer propagation of a piece of information from individual $i$ to $j$ is driven by two properties: $i$'s influence and $j$'s susceptibility. This assumption is motivated by the findings of a large-scale field experiment on new product adoption in Facebook~\cite{aral2012identifying}.
Specifically, the state of a given individual $i$ is characterized by a binary variable: she has either reshared (``retweeted") a piece of information ($r_i=1$) or not ($r_i=0$). The probability $p_{ij}$ that individual $j$ reshares a piece of content after $i$ resharing it (i.e., that the piece of information ``propagates" from $i$ to $j$) is defined as $p_{ij}=A_{ij}\,I_i\,S_j$, where $I_i$ and $S_j$ represent individual $i$'s influence and individual $j$'s susceptibility, respectively, and $A_{ij}$ is equal to one if $i$ and $j$ are connected in the diffusion network where information propagates~\cite{pei2014searching}, equal to zero otherwise. In this information diffusion network, a directed link from $i$ to $j$ exists if and only if at least one spreading event from $i$ to $j$ was recorded. This network might differ from the underlying structural network, where links exist based on the "follow" relationship between users, and it is better suited to analyze the spreading dynamics~\cite{pei2014searching}. 
Note that according to the model's terminology, a high-influence individual is effective at locally propagating information toward her social contacts, but might not be necessarily able to trigger large-scale propagation events.
Using the model to generate synthetic data on diffusion processes reveals that the diffusion success outcomes are highly dependent on the assortativity properties of influence and susceptibility~\cite{aral2018social}.

We consider here a different problem: Given empirical data on multiple spreading events, can we reliably reconstruct individuals' influence and susceptibility?
 Under the assumption of the diffusion model above, we derive two coupled non-linear equations that capture the relation between individual influence, individual susceptibility, and network structure. Solving iteratively the derived equations leads to an algorithm to estimate individual influence and susceptibility from multiple spreading event data (see Methods for the algorithm derivation). In the following, we refer to the scores determined by the algorithm as the \textit{reconstructed influence and susceptibility scores}, $\hat{I}_i$ and $\hat{S}_j$; we refer to this derived algorithm through which to obtain $\hat{I}_i$ and $\hat{S}_j$ as the \textit{influence-susceptibility (IS) algorithm}. The non-linear equations that define the algorithm indicate that when estimating individual-level influence from multiple spreading data, one needs to consider both the rate at which a node influences her peers (the higher, the higher the node's influence) and her peers' average susceptibility (the higher, the lower the node's influence; see Fig.~\ref{fig:illustration}a). Similarly, when estimating individual-level susceptibility from multiple spreading data, one needs to consider both the rate at which a node is influenced by her peers (the higher, the higher the node's susceptibility) and her peers' average influence (the higher, the lower the node's susceptibility; see Fig.~\ref{fig:illustration}b).

We validate the IS algorithm's outputs via two main findings. (\romannumeral1) First, we find that in synthetic spreading data generated with the empirical diffusion model, the IS algorithm can reconstruct almost perfectly the latent ``ground-truth" influence and susceptibility scores of the nodes (see Supplementary Note $1$). This finding is robust with respect to incomplete data: A high reconstruction accuracy can be maintained even when a substantial portion of the spreading events are removed (see Supplementary Note $2$).
 (\romannumeral2) Second, we find that in empirical spreading data from Twitter and Weibo, the algorithm produces individual-level scores that respect the four stylized facts about influence and susceptibility identified in 
 's experiment~\cite{aral2012identifying} (see Supplementary Note $3$). Specifically, our findings confirm the following four properties:
 \begin{itemize}
     \item Highly influential individuals tend not to be susceptible, highly susceptible individuals tend not to be influential, and almost no one is both highly influential and highly susceptible to influence.
     \item Both influential individuals and non-influential individuals have approximately the same distribution of susceptibility to influence among their peers.
     \item There are more people with high influence scores than high susceptibility scores.
     \item Influentials cluster in the network.
 \end{itemize}
We refer to Supplementary Note $3$ for the results that demonstrate that the reconstructed IS scores respect these properties. Taken together, these properties suggest that the scores calculated by the IS algorithm are consistent with the influence and susceptibility traits as defined experimentally by Aral and Walker~\cite{aral2012identifying}.

 \begin{figure*}[t]
    \centering

    \includegraphics[scale=0.36]{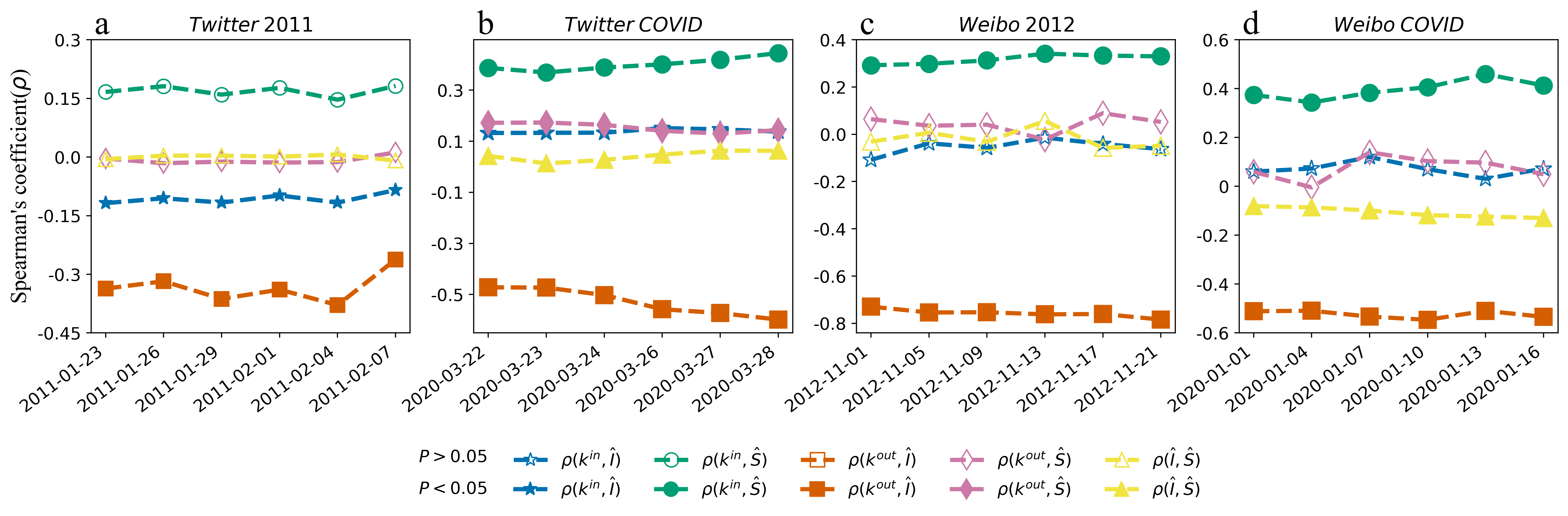}
\caption{{\bf Empirical correlations between individual-level properties.} We divide the spreading events into $6$ consecutive non-overlapping periods. The date on the $x$-axis represents the starting date of each period. In each period, we reconstruct individual influence and susceptibility via the IS algorithm, and measure the Spearman's correlation coefficients, $\rho$, between indegree ($k^{in}$), outdegree ($k^{out}$), influence ($\hat{I}$) and susceptibility ($\hat{S}$). Only those value pairs where both values are not equal to $0$ are conserved. Filled markers denote correlation values that are significantly larger or smaller than the correlation values calculated on randomized networks ($P<0.05$, see Supplementary Note $4$ for details); empty markers denote correlation values that do not significantly differ from the correlation values calculated on randomized networks ($P>0.05$). The strongest consistent empirical correlation is the one between outdegree ($k^{out}$) and influence ($\hat{I}$); the other correlations are weak or non-significant.}
    \label{fig:corr}
\end{figure*}

\begin{figure*}[t]
    \centering
    \includegraphics[scale=0.36]{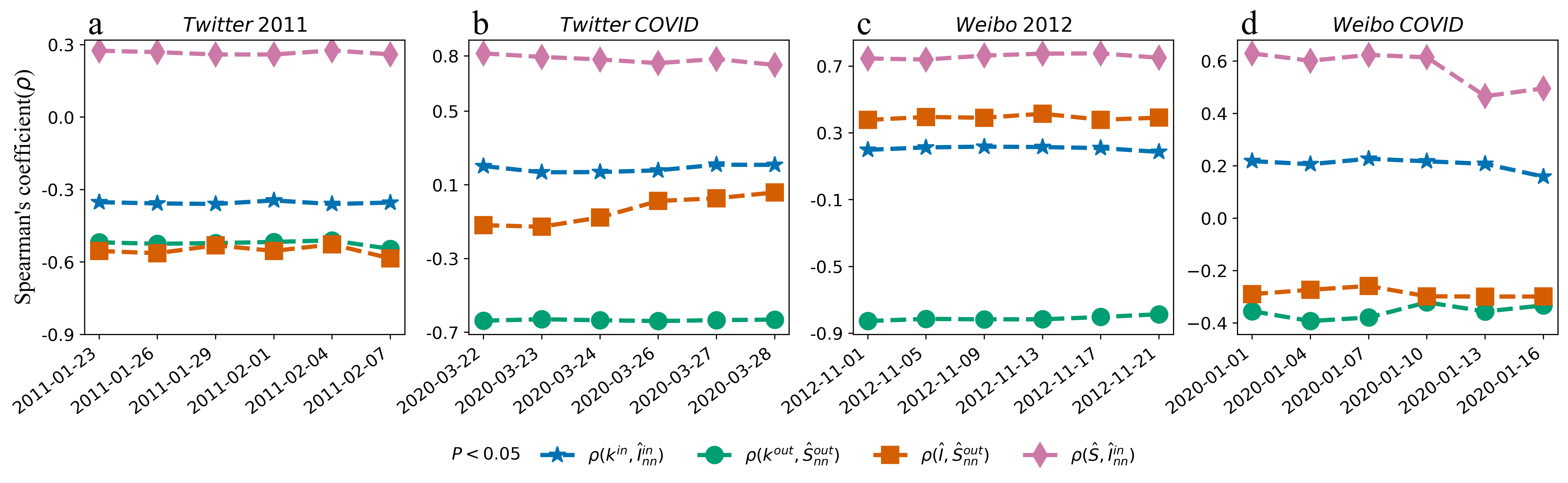}
   \caption{{\bf Empirical assortativity properties between degree, influence, and susceptibility.} We divide the spreading events into 6  consecutive non-overlapping periods. The date on the $x$-axis represents the starting date of each period. In each period, we reconstruct individual influence and susceptibility via the IS algorithm, and we measure Spearman's correlation coefficients, $\rho$, between an individual's properties and her neighbors' properties. The only consistent correlations are the positive one between an individual's susceptibility ($\hat{S}$) and the influence of the individuals she retweets ($\hat{I}_{nn}^{in}$), and the negative one between an individual's outdegree ($k^{out}$) and the susceptibility of the individuals who retweet from her ($\hat{S}_{nn}^{out}$).} 
    \label{fig:assortativity}
\end{figure*}

\textbf{The relation between individual influence, susceptibility, and degree.} We apply the IS algorithm to empirical spreading data to determine the relation between individuals' degree, influence, and susceptibility. Specifically, we examine: (\romannumeral1) the node-level correlation between an individual's properties, i.e., influence, susceptibility, and degree (see Fig.~\ref{fig:corr}; see Supplementary Fig.~S15 for additional results with the PageRank and $k$-core centralities); (\romannumeral2) the correlation between an individual's properties (influence, susceptibility, and degree) and her neighbors' properties (see Fig.~\ref{fig:assortativity}), which represents the assortativity of these node-level properties.
Revealing these empirical correlations is important because numerical simulations have shown that they can dramatically alter the outcome of spreading processes and the effectiveness of seeding strategies~\cite{aral2018social}.
To assess the statistical significance of the correlations between individuals' properties, we measure the $P$ values by comparing their empirical values against the values observed for randomized networks (see Supplementary Note $4$).

When examining node-level correlations, we observe that 
an individual's outdegree ($k^{out}$) and influence ($\hat{I}$) are the only pair of variables with a significant correlation across all four datasets.
Aligning with previous arguments~\cite{watts2007influentials,bakshy2011everyone}, the observed negative correlation suggests that individuals with many followers might not be the most effective ones at influencing their neighbors.
 At the same time, all other correlations are weak and dataset-dependent, suggesting that influence, susceptibility, and degree are orthogonal properties of individuals.
 

As for assortativity properties, we focus on four correlations between an individual's properties and her in- or out-neighbors properties: $\rho(k^{in},\hat{I}^{in}_{nn})$,   $\rho(k^{out},\hat{S}^{out}_{nn})$, $\rho(\hat{I},\hat{S}^{out}_{nn})$,   $\rho(\hat{S},\hat{I}^{in}_{nn})$;  in this notation, the subscript $nn$ indicates that we average the property over an individual's nearest neighbors, and the superscript $in/out$ indicates the in- or out-neighbors, respectively. 
Here we find two consistent correlations: the positive one between an individual's susceptibility ($\hat{S}$) and the influence of the individuals she is influenced by ($\hat{I}_{nn}^{in}$), and the negative one between an individual's outdegree ($k^{out}$) and the susceptibility of the individuals who retweet from her ($\hat{S}_{nn}^{out}$). The other correlations are inconsistent, meaning that their sign switches from positive to negative depending on the dataset. These observed correlations and similar observations shed light on the assortativity properties of degree, influence, and susceptibility, and they could serve as input for calibrated diffusion simulations to test the effectiveness of seeding programs~\cite{aral2018social}.

\begin{figure*}[t]
    \centering
    \includegraphics[scale=0.4]{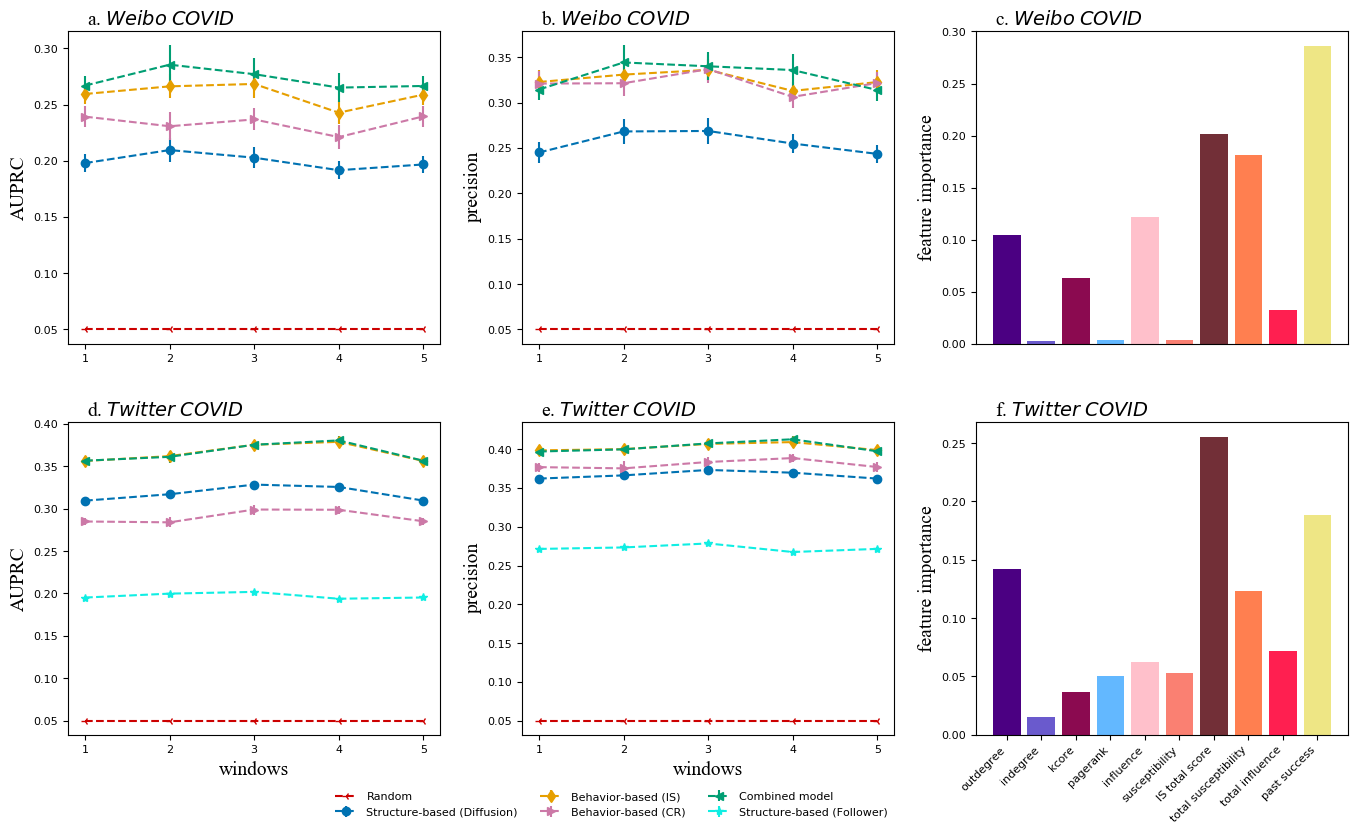}
    \caption{{\bf Predicting superspreaders.} We rely on the Random Forest classification algorithm and use different features as input to predict superspreaders, where the superspreaders are defined as individuals with top $5\%$ spreading capacity. Panels {\bf a, b, d, e} show the superspreaders predicting performance on \emph{Weibo COVID} and \emph{Twitter COVID}. Two metrics, AURPC, and precision, are adopted to measure the performance of models. Panels {\bf c, f} show the feature importance resulting from training the combined model.  Across all windows, the best-performing model is either the combined model or the Behavior-based (IS) model, which points to the essential role of the IS scores for the superspreader prediction. Panels {\bf c, f} show the feature importance obtained from training the combined model. IS-based features tend to be more important than centrality-based features. } 
    \label{fig:ssp} 
\end{figure*}

\textbf{Predicting superspreaders.} The above results indicate that the IS algorithm helps better understand the position of influential and susceptible nodes in empirical networks. Here we will show that it can be used to forecast who will be future superspreaders, namely individuals who can trigger large-scale information cascades.
To this end, we focus on the two large-scale datasets (\emph{Weibo COVID} and \emph{Twitter COVID}) for which the ID information of each spreading event is available, while we exclude the remaining datasets (\textit{Weibo 2012} and \textit{Twitter 2011}) that do not include this essential information.
We frame a predictive problem where, given data from a time window $t$, we seek to identify who will be the superspreaders in subsequent window $t+1$. More specifically, we split each of the two analyzed datasets (\emph{Weibo COVID} and \emph{Twitter COVID}) into a set of non-overlapping periods $\{d_1,d_2,\dots,d_T\}$ by the retweet time, and we use the IS scores and other metrics estimated in period $d_t$ to detect the superspreaders in period $d_{t+1}$. Superspreaders are defined as the top $z\%=5\%$ users by the average size of the cascades they initiate. To ensure the robustness of the results, we also implement additional selectivity thresholds of $1\%$, $2\%$, and $10\%$ respectively, and the corresponding results can be found in Supplementary Figs. S28--S30. 

In each training period $d_{t}$, for a given individual $i$, we consider various classes of predictors. Specifically, we consider five
new individual-level predictors based on the IS algorithm: 
(\romannumeral1) $i$'s \textit{influence}, $I_{i}=\hat{I}_{i;t}$; 
(\romannumeral2) $i$'s \textit{susceptibility}, $S_{i}=\hat{S}_{i;t}$; 
(\romannumeral3) the \textit{total susceptibility} of the individuals influenced by $i$; $TS_{i}=\sum_j A_{ij;t}\, \hat{S}_{j;t}$; (\romannumeral4) the \textit{total influence} of the individuals influenced by $i$; $TI_{i}=\sum_{j;t} A_{ij;t}\, \hat{I}_{j;t}$; (\romannumeral5) the \textit{IS total score} that information spreads from $i$ to her neighbors; $TIS_{i}=\sum_j A_{ij;t} \hat{{I}_{i;t}}\,\hat{S}_{j;t}$, motivated by the empirical influence model.
To illustrate the advantage of the IS-based predictors compared to simpler behavior-based predictors, we also consider an individual's total \textit{in-contagion rate}
and \textit{out-contagion rate}. The contagion rate from $i$ to $j$ at period $d_{t}$ (i.e., the fraction of $i$'s reshared pieces of information that are also subsequently reshared by $j$ within time window $d_{t}$) is denoted as  $\omega_{ij;t}$.  $j$'s in-contagion rate is defined as ${\hat{g_{j;t}}}=\sum_{i} A_{ij;t}\, \omega_{ij;t}$; $i$'s out-contagion rate is defined as ${\hat{f_{i;t}}}=\sum_{j} A_{ij;t}\, \omega_{ij;t}$.  
Motivated by the vast literature on social hubs and superspreaders~\cite{goldenberg2009role,pei2014searching, wang2019anomalous, lu2016vital}, we consider two network-based metrics as well: the outdegree, $k_{i;t}^{out}$ (reflecting the node's total number of influenced neighbors), and the indegree, $k_{i;t}^{in}$  (reflecting the node's total number of nodes that influenced $i$). 
Given the previously-documented effectiveness of past success in predicting future success for social media users~\cite{bakshy2011everyone,martin2016exploring}, we also consider the average size of user $i$'s previously-initiated cascades, which we refer to as past success.

To understand the role of the above predictors in the superspreader prediction problem, we consider various classes of predictive models (see Supplementary Note $8$ for details), which differ in the types of features they include: Two behavior-based models, with one based on the 4 IS-based features [Behavior-based (IS) model], while the other based on the contagion rate metrics [Behavior-based (CR) model];  A structure-based model that includes the indegree, outdegree, PageRank, and $k$-core centrality metrics extracted from the diffusion graph; A combined model that includes all the features in the behavior-based and structure-based models as well as the users' past success.
We consider a random classifier as a baseline model.
Finally, in \emph{Twitter COVID}, we are also able to consider a structure-based model based on the nodes' indegree and outdegree in the follower-followee graph.
We refer to Table~S4 in Supplementary Material for a summary of the features included in all the considered models. 

We train different machine-learning algorithms: the Random Forest~\cite{breiman2001random} (results shown in the main text), XGBoost algorithm~\cite{chen2016xgboost}, and Multi-Layer Perceptron~\cite{gardner1998artificial}, and results for the XGBoost and MLP are shown in Supplementary Figs. S31--S36.
We evaluate the models according to two out-of-sample metrics: the Area Under the Precision-Recall Curve (AUPRC), which is better suited than the popular AUC metric to evaluate classifier performance for imbalanced problems~\cite{davis2006relationship,saito2015precision}; the precision of the metrics in identifying the superspreaders.
We refer to Supplementary Note $8$ for all the details of the models' parameters setting and training.

We find that across all windows, the best-performing model is either the combined model or the Behavior-based (IS) model (Figs.~\ref{fig:ssp}a--d). 
The Behavior-based (IS) model outperforms the structural models, which points to the stronger predictive signals in the behavior-based features compared to structural ones. Performance differences between the combined model and the Behavior-based (IS) model are only observed in Weibo but not in Twitter, and they tend to be relatively small. This indicates that adding the past success feature to the information already included in the behavior-based features brings no substantial accuracy gains.
Finally, we note that the Behavior-based (IS) model tends to substantially outperform the Behavior-based (CR) model (Figs.~\ref{fig:ssp}a--d), which points to the superior predictive power of the IS algorithm scores compared to the contagion rates. While these conclusions were obtained through the Random Forest algorithm, to a large extent, they also hold for the XGBoost and MLP algorithms (see Supplementary Figs. ~S31--S36), which indicates their robustness across different learning algorithms. 

 Additional insights into the position of superspreaders can be gained by looking at feature importance in the combined model trained with the Random Forest algorithm (Figs.~\ref{fig:ssp}e--h). The features based on the IS scores tend to be more important than the structure-based features. The IS total score of a node is consistently among the two most important predictors; only the past success is a more important feature in Weibo. These findings are robust with respect to different thresholds for the superspreader detection (e.g., top-$1\%$, top-$2\%$, and top-$10\%$, see Supplementary Figs. ~S28--S30). 

To better characterize the superspreaders' network position, we perform a linear regression analysis with a linear mixed-effect model. 
After removing a multicollinearity issue (see Supplementary Note $7$ for details), we find that the IS total score, the total influence, and the $k$-core centrality~\cite{pei2014searching} are the only variables with a significant association with the users' spreading success in both Twitter and Weibo COVID (see Supplementary Note $7$, Tables~S17, S18 for details). 
The found positive effect of the IS total score suggests that the superspreaders are characterized by more high-contagion links (i.e., directed links with high $\hat{I}\times\hat{S}$); the positive effect of the total influence suggests that they tend to influence more influential users.
Taken together, these results indicate that behavior-based variables based on the IS algorithm can substantially improve the accuracy of superspreader predictions above and beyond models network centrality, and reveal new insights on the network position of the superspreaders.

\section*{CONCLUSION}

Prior works on the role of individuals in online information diffusion focused either on individuals' network centrality or their behavioral traits such as influence and susceptibility. However, these studies did not clarify whether individuals' network centrality or behavioral traits are better predictors of diffusion success.
In this work, we propose an iterative algorithm that can reliably infer influence and susceptibility from empirical data on multiple spreading events in social media.
Therefore, unlike previous studies that merely relied on parsimonious assumptions to explore the micro-level mechanism behind the spreading process, our algorithm provides reliable empirical scores of network individuals, allowing for a more precise understanding of the information diffusion process.
In general, the proposed approach offers more accurate predictions of future superspreaders compared to methods that only rely on network-based structural characteristics.
Notably, the proposed algorithm simultaneously incorporates information on the global network structure and peer-to-peer propagation events. Differently from previous studies that emphasized the role of highly-connected social hubs~\cite{goldenberg2009role,pei2014searching, wang2019anomalous, lu2016vital} or highly-influential individuals~\cite{wang2019anomalous}, our results indicate that to understand and accurately predict information spreading in social media, inferring both individuals' influence and susceptibility is necessary.

Our work opens exciting research directions.
First, we have focused on social platforms (Twitter and Weibo) where users reshare content from other users by explicitly mentioning them. When such an explicit mention is absent, it is challenging to distinguish between content propagation from user to user and other possible reasons for sharing, like advertising and general media influence~\cite{pei2014searching}.
Besides, generalizing our algorithm to quantify individual influence and susceptibility from generic adoption time-series would require careful consideration to factor out potential homophily effects, which notoriously lead to an overestimation of social contagion estimates~\cite{aral2009distinguishing}.  Furthermore, beyond social contagion, our method may prove useful in identifying influential and susceptible individuals in epidemic spreading processes, where ``superspreaders'' can infect disproportionately many other individuals~\cite{lloyd2005superspreading}, whereas some individual groups might be significantly more vulnerable to the severe consequences of a disease~\cite{godri2020covid}. In general, our algorithm may prove useful wherever there is peer-to-peer contagion, and there is heterogeneity in individual units' influence and susceptibility; it can also shed some light on the network dismantling and population immunization problems~\cite{lu2016vital,liu2021efficient}.

To conclude, our findings could have implications for behavioral-change campaigns, such as viral marketing campaigns or community-minded behavioral nudging. These programs traditionally focus on the most central individuals in a given social network~\cite{hinz2011seeding, banerjee2013diffusion,lu2016vital,muller2019effect}. However, our results suggest that better strategies might involve identifying the network connections between highly influential and highly susceptible individuals. Given that our findings rely on predictive insights, additional research is needed in order to generalize them and develop intervention strategies. For instance, future studies may leverage observational data and field experiments, exploring how to best integrate network and behavioral information to design policies aimed at large-scale behavioral change.

\section*{METHODS}


\textbf{Influence-susceptibility algorithm derivation and convergence.} 
The main idea of the derivation is to obtain an equation for the individual-level scores by summing $p_{ij}$ over all possible source nodes $i$ (to derive an equation for $j$'s susceptibility) and over all possible target nodes $j$ (to derive an equation for $i$'s influence).
For each individual $i$, we sum the contagion probability $p_{ij}$ over all $i$'s neighbors:
\begin{equation}
    \sum_{j} p_{ij}=\sum_{j}A_{ij}\, p_{ij} = I_i\,\Bigl(\sum_j A_{ij} S_j\Bigr).
    \label{eq1}
\end{equation}
 Similarly, for each individual $j$, we sum the contagion probability $p_{ij}$ over all her neighbors that can influence her:
\begin{equation}
    \sum_{i} p_{ij}= \sum_{i} A_{ij}\, p_{ij}=\Bigl(\sum_i A_{ij} I_i\Bigr)\, S_j.
    \label{eq2}
\end{equation}
For convenience, we define $f_i=\sum_j A_{ij}\,p_{ij}$ and $g_j=\sum_i A_{ij}\,p_{ij}$; these two variables can be interpreted as $i$'s outgoing contagion rate (i.e., how frequently $i$'s neighbors reshare the pieces of information shared by $i$) and $j$'s incoming contagion rate, respectively (i.e., how frequently $j$ reshares the pieces of information shared by her neighbors). The rates $f_i$ and $g_j$ can be interpreted as rough estimates of $i$'s influence on her neighbors and $j$'s susceptibility to her neighbors, respectively.
We can rewrite equation~\eqref{eq1} as
\begin{equation}
   I_i= \frac{f_i}{\sum_j A_{ij} S_j}.
   \label{eq3}
\end{equation}
Similarly, from equation~\eqref{eq2}, we obtain
\begin{equation}
     S_j=\frac{g_j}{\sum_i A_{ij} I_i}.
    \label{eq4}
\end{equation}
The set~\eqref{eq3}--\eqref{eq4} of equations indicate that to infer influence and susceptibility from observational data, a nonlinear algorithm is needed. The nonlinearity of equation~\eqref{eq3} stems from the fact that given data on multiple spreading events, an individual $i$ is ``highly-influential" (large $I_i$) if she has a large outgoing contagion rate $f_i$, and low
$\sum_j A_{ij}\,S_j$, indicating that overall, individuals who are influenced by $i$ are not highly susceptible. 
Similarly, from equation~\eqref{eq4}, an individual $j$ is ``highly-susceptible" (large $S_j$) if she has a large incoming contagion rate $g_j$, and low
$\sum_i A_{ij}\,I_i$, indicating that overall, individuals who influence $j$ are not highly influential.
In Fig.~\ref{fig:illustration}, we provide an illustration of the interpretation of the influence $I$ and susceptibility $S$ by equations (\ref{eq3})-(\ref{eq4}). 

Leveraging these two equations to estimate individual influence and susceptibility faces two potential obstacles. First, to estimate $f$ and $g$, we need to know the contagion probability $p_{ij}$, which is unknown a priori. Second, estimating the scores $I_i$ and $S_j$ of individual $i$ requires knowledge of her neighbors' susceptibility score and influence score, respectively.

To estimate $f$ and $g$, we infer $p_{ij}$ from the spreading data by measuring the fraction of $i$'s reshared pieces of information that are also subsequently reshared by $j$, $\omega_{ij}$, and thus produce estimates $\hat{f}_i=\sum_j A_{ij}\,\omega_{ij}$ and $\hat{g}_j=\sum_i A_{ij}\,\omega_{ij}$ of the outgoing and incoming contagion rate, respectively.
To overcome the second obstacle, inspired by state-of-the-art economic complexity algorithms~\cite{tacchella2012new}, we estimate the scores iteratively through the following nonlinear map
\begin{equation}
\begin{split}
   \hat{I}_i^{(n+1)}&= \frac{\hat{f}_i}{\sum_j A_{ij} \hat{S}_j^{(n)}},\\
     \hat{S}_j^{(n+1)} &=\frac{\hat{g}_j}{\sum_i A_{ij} \hat{I}_i^{(n)}},
    \label{map}
    \end{split}
\end{equation}
By appropriately setting the initial condition, the convergence of this algorithm is guaranteed, and the detailed proofs and conclusions refer to Supplementary Note $5$. For further discussion and extension of the IS algorithm please see Supplementary Note $10$.

%

\section*{DATA AVAILABILITY}
The data collection, sampling, and partition for the $4$ large-scale empirical datasets are described in Supplementary Note $9$.
The source code of the IS algorithm is available at \href{https://github.com/zervel3/IS-algorithm}{link}.

\section*{FUNDING}

 LL acknowledges support from the National Natural Science Foundation of China (T2293771), the STI 2030—Major Projects (2022ZD0211400), the Sichuan Province Outstanding Young Scientists Foundation (2023NSFSC1919), and the New Cornerstone Science Foundation through the XPLORER PRIZE. JGL acknowledges support from the National Natural Science Foundation of China (72371150, 72032003), and the Fundamental Research Funds for the Central Universities: High-Quality Development 23 of Digital Economy: An Investigation of Characteristics and Driving Strategies (2023110139). MSM acknowledges financial support from the Swiss National Science Foundation (100013-207888) and the URPP Social Networks at the University of Zurich.

\section*{AUTHOR CONTRIBUTIONS}

	LL and MSM conceived and designed the experiments; FZ and JGL collected the empirical datasets; FZ and MSM developed the new algorithm; FZ performed the numeric analysis; all the authors analyzed the data; FZ and MSM wrote the first draft of the manuscript; all the authors revised the manuscript.

{\bf Conflict of interest statement.} The authors declare no competing interests.


\bibliography{reference}

\begin{thebibliography}{10}
\expandafter\ifx\csname url\endcsname\relax
  \def\url#1{\texttt{#1}}\fi
\expandafter\ifx\csname urlprefix\endcsname\relax\def\urlprefix{URL }\fi
\providecommand{\bibinfo}[2]{#2}
\providecommand{\eprint}[2][]{\url{#2}}

\bibitem{bak2021stewardship}
\bibinfo{author}{Bak-Coleman, J.~B.} \emph{et~al.}
\newblock \bibinfo{title}{Stewardship of global collective behavior}.
\newblock \emph{\bibinfo{journal}{Proc Natl Acad Sci U S A}} \textbf{\bibinfo{volume}{118}} (\bibinfo{year}{2021}).

\bibitem{lacassin2020rapid}
\bibinfo{author}{Lacassin, R.} \emph{et~al.}
\newblock \bibinfo{title}{Rapid collaborative knowledge building via twitter after significant geohazard events}.
\newblock \emph{\bibinfo{journal}{Geosci Lett}} \textbf{\bibinfo{volume}{3}}, \bibinfo{pages}{129--146} (\bibinfo{year}{2020}).

\bibitem{chen2010social}
\bibinfo{author}{Chen, Y.}, \bibinfo{author}{Harper, F.~M.}, \bibinfo{author}{Konstan, J.} \& \bibinfo{author}{Li, S.~X.}
\newblock \bibinfo{title}{Social comparisons and contributions to online communities: A field experiment on movielens}.
\newblock \emph{\bibinfo{journal}{Am Econ Rev}} \textbf{\bibinfo{volume}{100}}, \bibinfo{pages}{1358--98} (\bibinfo{year}{2010}).

\bibitem{centola2018behavior}
\bibinfo{author}{Centola, D.}
\newblock \emph{\bibinfo{title}{How behavior spreads}} (\bibinfo{publisher}{New Jersey: Princeton University Press}, \bibinfo{year}{2018}).

\bibitem{huber2019fostering}
\bibinfo{author}{Huber, B.}, \bibinfo{author}{Barnidge, M.}, \bibinfo{author}{Gil~de Z{\'u}{\~n}iga, H.} \& \bibinfo{author}{Liu, J.}
\newblock \bibinfo{title}{Fostering public trust in science: The role of social media}.
\newblock \emph{\bibinfo{journal}{Public Underst Sci}} \textbf{\bibinfo{volume}{28}}, \bibinfo{pages}{759--777} (\bibinfo{year}{2019}).

\bibitem{lazer2018science}
\bibinfo{author}{Lazer, D.~M.} \emph{et~al.}
\newblock \bibinfo{title}{The science of fake news}.
\newblock \emph{\bibinfo{journal}{Science}} \textbf{\bibinfo{volume}{359}}, \bibinfo{pages}{1094--1096} (\bibinfo{year}{2018}).

\bibitem{shao2018spread}
\bibinfo{author}{Shao, C.} \emph{et~al.}
\newblock \bibinfo{title}{The spread of low-credibility content by social bots}.
\newblock \emph{\bibinfo{journal}{Nat Commun}} \textbf{\bibinfo{volume}{9}}, \bibinfo{pages}{1--9} (\bibinfo{year}{2018}).

\bibitem{bovet2019influence}
\bibinfo{author}{Bovet, A.} \& \bibinfo{author}{Makse, H.~A.}
\newblock \bibinfo{title}{Influence of fake news in twitter during the 2016 us presidential election}.
\newblock \emph{\bibinfo{journal}{Nat Commun}} \textbf{\bibinfo{volume}{10}}, \bibinfo{pages}{1--14} (\bibinfo{year}{2019}).

\bibitem{guess2020exposure}
\bibinfo{author}{Guess, A.~M.}, \bibinfo{author}{Nyhan, B.} \& \bibinfo{author}{Reifler, J.}
\newblock \bibinfo{title}{Exposure to untrustworthy websites in the 2016 us election}.
\newblock \emph{\bibinfo{journal}{Nat Hum Behav}} \textbf{\bibinfo{volume}{4}}, \bibinfo{pages}{472--480} (\bibinfo{year}{2020}).

\bibitem{gallotti2020assessing}
\bibinfo{author}{Gallotti, R.}, \bibinfo{author}{Valle, F.}, \bibinfo{author}{Castaldo, N.}, \bibinfo{author}{Sacco, P.} \& \bibinfo{author}{De~Domenico, M.}
\newblock \bibinfo{title}{Assessing the risks of ‘infodemics’ in response to covid-19 epidemics}.
\newblock \emph{\bibinfo{journal}{Nat Hum Behav}} \textbf{\bibinfo{volume}{4}}, \bibinfo{pages}{1285--1293} (\bibinfo{year}{2020}).

\bibitem{stella2018bots}
\bibinfo{author}{Stella, M.}, \bibinfo{author}{Ferrara, E.} \& \bibinfo{author}{De~Domenico, M.}
\newblock \bibinfo{title}{Bots increase exposure to negative and inflammatory content in online social systems}.
\newblock \emph{\bibinfo{journal}{Proc Natl Acad Sci U S A}} \textbf{\bibinfo{volume}{115}}, \bibinfo{pages}{12435--12440} (\bibinfo{year}{2018}).

\bibitem{johnson2020online}
\bibinfo{author}{Johnson, N.~F.} \emph{et~al.}
\newblock \bibinfo{title}{The online competition between pro-and anti-vaccination views}.
\newblock \emph{\bibinfo{journal}{Nature}} \textbf{\bibinfo{volume}{582}}, \bibinfo{pages}{230--233} (\bibinfo{year}{2020}).

\bibitem{baumann2020modeling}
\bibinfo{author}{Baumann, F.}, \bibinfo{author}{Lorenz-Spreen, P.}, \bibinfo{author}{Sokolov, I.~M.} \& \bibinfo{author}{Starnini, M.}
\newblock \bibinfo{title}{Modeling echo chambers and polarization dynamics in social networks}.
\newblock \emph{\bibinfo{journal}{Phys Rev Lett}} \textbf{\bibinfo{volume}{124}}, \bibinfo{pages}{048301} (\bibinfo{year}{2020}).

\bibitem{medo2021fragility}
\bibinfo{author}{Medo, M.}, \bibinfo{author}{Mariani, M.~S.} \& \bibinfo{author}{L{\"u}, L.}
\newblock \bibinfo{title}{The fragility of opinion formation in a complex world}.
\newblock \emph{\bibinfo{journal}{Commun Phys}} \textbf{\bibinfo{volume}{4}}, \bibinfo{pages}{1--10} (\bibinfo{year}{2021}).

\bibitem{gonzalez2011dynamics}
\bibinfo{author}{Gonz{\'a}lez-Bail{\'o}n, S.}, \bibinfo{author}{Borge-Holthoefer, J.}, \bibinfo{author}{Rivero, A.} \& \bibinfo{author}{Moreno, Y.}
\newblock \bibinfo{title}{The dynamics of protest recruitment through an online network}.
\newblock \emph{\bibinfo{journal}{Sci Rep}} \textbf{\bibinfo{volume}{1}}, \bibinfo{pages}{1--7} (\bibinfo{year}{2011}).

\bibitem{mooijman2018moralization}
\bibinfo{author}{Mooijman, M.}, \bibinfo{author}{Hoover, J.}, \bibinfo{author}{Lin, Y.}, \bibinfo{author}{Ji, H.} \& \bibinfo{author}{Dehghani, M.}
\newblock \bibinfo{title}{Moralization in social networks and the emergence of violence during protests}.
\newblock \emph{\bibinfo{journal}{Nat Hum Behav}} \textbf{\bibinfo{volume}{2}}, \bibinfo{pages}{389--396} (\bibinfo{year}{2018}).

\bibitem{aral2018social}
\bibinfo{author}{Aral, S.} \& \bibinfo{author}{Dhillon, P.~S.}
\newblock \bibinfo{title}{Social influence maximization under empirical influence models}.
\newblock \emph{\bibinfo{journal}{Nat Hum Behav}} \textbf{\bibinfo{volume}{2}}, \bibinfo{pages}{375--382} (\bibinfo{year}{2018}).

\bibitem{hu2018local}
\bibinfo{author}{Hu, Y.} \emph{et~al.}
\newblock \bibinfo{title}{Local structure can identify and quantify influential global spreaders in large scale social networks}.
\newblock \emph{\bibinfo{journal}{Proc Natl Acad Sci U S A}} \textbf{\bibinfo{volume}{115}}, \bibinfo{pages}{7468--7472} (\bibinfo{year}{2018}).

\bibitem{wang2019anomalous}
\bibinfo{author}{Wang, X.}, \bibinfo{author}{Lan, Y.} \& \bibinfo{author}{Xiao, J.}
\newblock \bibinfo{title}{Anomalous structure and dynamics in news diffusion among heterogeneous individuals}.
\newblock \emph{\bibinfo{journal}{Nat Hum Behav}} \textbf{\bibinfo{volume}{3}}, \bibinfo{pages}{709--718} (\bibinfo{year}{2019}).

\bibitem{shi2019totally}
\bibinfo{author}{Shi, D.}, \bibinfo{author}{L{\"u}, L.} \& \bibinfo{author}{Chen, G.}
\newblock \bibinfo{title}{Totally homogeneous networks}.
\newblock \emph{\bibinfo{journal}{Natl Sci Rev}} \textbf{\bibinfo{volume}{6}}, \bibinfo{pages}{962--969} (\bibinfo{year}{2019}).

\bibitem{tang2020predictability}
\bibinfo{author}{Tang, D.} \emph{et~al.}
\newblock \bibinfo{title}{Predictability of real temporal networks}.
\newblock \emph{\bibinfo{journal}{Natl Sci Rev}} \textbf{\bibinfo{volume}{7}}, \bibinfo{pages}{929--937} (\bibinfo{year}{2020}).

\bibitem{xie2021detecting}
\bibinfo{author}{Xie, J.} \emph{et~al.}
\newblock \bibinfo{title}{Detecting and modelling real percolation and phase transitions of information on social media}.
\newblock \emph{\bibinfo{journal}{Nat Hum Behav}} \bibinfo{pages}{1--8} (\bibinfo{year}{2021}).

\bibitem{zhou2020realistic}
\bibinfo{author}{Zhou, B.} \emph{et~al.}
\newblock \bibinfo{title}{Realistic modelling of information spread using peer-to-peer diffusion patterns}.
\newblock \emph{\bibinfo{journal}{Nat Hum Behav}} \textbf{\bibinfo{volume}{4}}, \bibinfo{pages}{1198--1207} (\bibinfo{year}{2020}).

\bibitem{juul2021comparing}
\bibinfo{author}{Juul, J.~L.} \& \bibinfo{author}{Ugander, J.}
\newblock \bibinfo{title}{Comparing information diffusion mechanisms by matching on cascade size}.
\newblock \emph{\bibinfo{journal}{Proc Natl Acad Sci U S A}} \textbf{\bibinfo{volume}{118}} (\bibinfo{year}{2021}).

\bibitem{hinz2011seeding}
\bibinfo{author}{Hinz, O.}, \bibinfo{author}{Skiera, B.}, \bibinfo{author}{Barrot, C.} \& \bibinfo{author}{Becker, J.~U.}
\newblock \bibinfo{title}{Seeding strategies for viral marketing: An empirical comparison}.
\newblock \emph{\bibinfo{journal}{Journal of Marketing}} \textbf{\bibinfo{volume}{75}}, \bibinfo{pages}{55--71} (\bibinfo{year}{2011}).

\bibitem{muller2019effect}
\bibinfo{author}{Muller, E.} \& \bibinfo{author}{Peres, R.}
\newblock \bibinfo{title}{The effect of social networks structure on innovation performance: A review and directions for research}.
\newblock \emph{\bibinfo{journal}{Int J Res Mark}} \textbf{\bibinfo{volume}{36}}, \bibinfo{pages}{3--19} (\bibinfo{year}{2019}).

\bibitem{budak2011limiting}
\bibinfo{author}{Budak, C.}, \bibinfo{author}{Agrawal, D.} \& \bibinfo{author}{El~Abbadi, A.}
\newblock \bibinfo{title}{Limiting the spread of misinformation in social networks}.
\newblock In \emph{\bibinfo{booktitle}{In: Proceedings of the 20th International Conference on World Wide Web}}, \bibinfo{pages}{665--674} (\bibinfo{year}{2011}).

\bibitem{grinberg2019fake}
\bibinfo{author}{Grinberg, N.}, \bibinfo{author}{Joseph, K.}, \bibinfo{author}{Friedland, L.}, \bibinfo{author}{Swire-Thompson, B.} \& \bibinfo{author}{Lazer, D.}
\newblock \bibinfo{title}{Fake news on twitter during the 2016 us presidential election}.
\newblock \emph{\bibinfo{journal}{Science}} \textbf{\bibinfo{volume}{363}}, \bibinfo{pages}{374--378} (\bibinfo{year}{2019}).

\bibitem{katz1955personal}
\bibinfo{author}{Katz, E.} \& \bibinfo{author}{Lazarsfeld, P.~F.}
\newblock \emph{\bibinfo{title}{Personal influence}} (\bibinfo{publisher}{New Jersey: Transaction Publishers}, \bibinfo{year}{1955}).

\bibitem{rogers2010diffusion}
\bibinfo{author}{Rogers, E.~M.}
\newblock \emph{\bibinfo{title}{Diffusion of innovations}} (\bibinfo{publisher}{Simon and Schuster}, \bibinfo{year}{2010}).

\bibitem{goldenberg2009role}
\bibinfo{author}{Goldenberg, J.}, \bibinfo{author}{Han, S.}, \bibinfo{author}{Lehmann, D.~R.} \& \bibinfo{author}{Hong, J.~W.}
\newblock \bibinfo{title}{The role of hubs in the adoption process}.
\newblock \emph{\bibinfo{journal}{J Mark}} \textbf{\bibinfo{volume}{73}}, \bibinfo{pages}{1--13} (\bibinfo{year}{2009}).

\bibitem{domingos2001mining}
\bibinfo{author}{Domingos, P.} \& \bibinfo{author}{Richardson, M.}
\newblock \bibinfo{title}{Mining the network value of customers}.
\newblock In \emph{\bibinfo{booktitle}{In: Proceedings of the Seventh ACM SIGKDD International Conference on Knowledge Discovery and Data Mining}}, \bibinfo{pages}{57--66} (\bibinfo{year}{2001}).

\bibitem{kempe2003maximizing}
\bibinfo{author}{Kempe, D.}, \bibinfo{author}{Kleinberg, J.} \& \bibinfo{author}{Tardos, {\'E}.}
\newblock \bibinfo{title}{Maximizing the spread of influence through a social network}.
\newblock In \emph{\bibinfo{booktitle}{In: Proceedings of the Ninth ACM SIGKDD International Conference on Knowledge Discovery and Data Mining}}, \bibinfo{pages}{137--146} (\bibinfo{year}{2003}).

\bibitem{kitsak2010identification}
\bibinfo{author}{Kitsak, M.} \emph{et~al.}
\newblock \bibinfo{title}{Identification of influential spreaders in complex networks}.
\newblock \emph{\bibinfo{journal}{Nat Phys}} \textbf{\bibinfo{volume}{6}}, \bibinfo{pages}{888--893} (\bibinfo{year}{2010}).

\bibitem{pei2014searching}
\bibinfo{author}{Pei, S.}, \bibinfo{author}{Muchnik, L.}, \bibinfo{author}{Andrade~Jr, J.~S.}, \bibinfo{author}{Zheng, Z.} \& \bibinfo{author}{Makse, H.~A.}
\newblock \bibinfo{title}{Searching for superspreaders of information in real-world social media}.
\newblock \emph{\bibinfo{journal}{Sci Rep}} \textbf{\bibinfo{volume}{4}}, \bibinfo{pages}{1--12} (\bibinfo{year}{2014}).

\bibitem{lu2016vital}
\bibinfo{author}{L{\"u}, L.} \emph{et~al.}
\newblock \bibinfo{title}{Vital nodes identification in complex networks}.
\newblock \emph{\bibinfo{journal}{Phys Rep-Rev Sec Phys Lett}} \textbf{\bibinfo{volume}{650}}, \bibinfo{pages}{1--63} (\bibinfo{year}{2016}).

\bibitem{zhou2019fast}
\bibinfo{author}{Zhou, F.}, \bibinfo{author}{L{\"u}, L.} \& \bibinfo{author}{Mariani, M.~S.}
\newblock \bibinfo{title}{Fast influencers in complex networks}.
\newblock \emph{\bibinfo{journal}{Commun Nonlinear Sci Numer Simul}} \textbf{\bibinfo{volume}{74}}, \bibinfo{pages}{69--83} (\bibinfo{year}{2019}).

\bibitem{banerjee2013diffusion}
\bibinfo{author}{Banerjee, A.}, \bibinfo{author}{Chandrasekhar, A.~G.}, \bibinfo{author}{Duflo, E.} \& \bibinfo{author}{Jackson, M.~O.}
\newblock \bibinfo{title}{The diffusion of microfinance}.
\newblock \emph{\bibinfo{journal}{Science}} \textbf{\bibinfo{volume}{341}}, \bibinfo{pages}{1236498} (\bibinfo{year}{2013}).

\bibitem{watts2007influentials}
\bibinfo{author}{Watts, D.~J.} \& \bibinfo{author}{Dodds, P.~S.}
\newblock \bibinfo{title}{Influentials, networks, and public opinion formation}.
\newblock \emph{\bibinfo{journal}{J Consum Res}} \textbf{\bibinfo{volume}{34}}, \bibinfo{pages}{441--458} (\bibinfo{year}{2007}).

\bibitem{galeotti2009influencing}
\bibinfo{author}{Galeotti, A.} \& \bibinfo{author}{Goyal, S.}
\newblock \bibinfo{title}{Influencing the influencers: a theory of strategic diffusion}.
\newblock \emph{\bibinfo{journal}{RAND J Econ}} \textbf{\bibinfo{volume}{40}}, \bibinfo{pages}{509--532} (\bibinfo{year}{2009}).

\bibitem{mariani2020wisdom}
\bibinfo{author}{Mariani, M.~S.} \emph{et~al.}
\newblock \bibinfo{title}{The wisdom of the few: Predicting collective success from individual behavior}.
\newblock \emph{\bibinfo{journal}{arXiv:2001.04777}}  (\bibinfo{year}{2020}).

\bibitem{rossman2021network}
\bibinfo{author}{Rossman, G.} \& \bibinfo{author}{Fisher, J.~C.}
\newblock \bibinfo{title}{Network hubs cease to be influential in the presence of low levels of advertising}.
\newblock \emph{\bibinfo{journal}{Proc Natl Acad Sci U S A}} \textbf{\bibinfo{volume}{118}}, \bibinfo{pages}{e2013391118} (\bibinfo{year}{2021}).

\bibitem{aral2012identifying}
\bibinfo{author}{Aral, S.} \& \bibinfo{author}{Walker, D.}
\newblock \bibinfo{title}{Identifying influential and susceptible members of social networks}.
\newblock \emph{\bibinfo{journal}{Science}} \textbf{\bibinfo{volume}{337}}, \bibinfo{pages}{337--341} (\bibinfo{year}{2012}).

\bibitem{bakshy2011everyone}
\bibinfo{author}{Bakshy, E.}, \bibinfo{author}{Hofman, J.~M.}, \bibinfo{author}{Mason, W.~A.} \& \bibinfo{author}{Watts, D.~J.}
\newblock \bibinfo{title}{Everyone's an influencer: quantifying influence on twitter}.
\newblock In \emph{\bibinfo{booktitle}{In: Proceedings of the fourth ACM International Conference on Web Search and Data Mining}}, \bibinfo{pages}{65--74} (\bibinfo{year}{2011}).

\bibitem{martin2016exploring}
\bibinfo{author}{Martin, T.}, \bibinfo{author}{Hofman, J.~M.}, \bibinfo{author}{Sharma, A.}, \bibinfo{author}{Anderson, A.} \& \bibinfo{author}{Watts, D.~J.}
\newblock \bibinfo{title}{Exploring limits to prediction in complex social systems}.
\newblock In \emph{\bibinfo{booktitle}{In: Proceedings of the 25th International Conference on World Wide Web}}, \bibinfo{pages}{683--694} (\bibinfo{year}{2016}).

\bibitem{breiman2001random}
\bibinfo{author}{Breiman, L.}
\newblock \bibinfo{title}{Random forests}.
\newblock \emph{\bibinfo{journal}{Mach Learn}} \textbf{\bibinfo{volume}{45}}, \bibinfo{pages}{5--32} (\bibinfo{year}{2001}).

\bibitem{chen2016xgboost}
\bibinfo{author}{Chen, T.} \& \bibinfo{author}{Guestrin, C.}
\newblock \bibinfo{title}{Xgboost: A scalable tree boosting system}.
\newblock In \emph{\bibinfo{booktitle}{Proceedings of the 22nd acm sigkdd international conference on knowledge discovery and data mining}}, \bibinfo{pages}{785--794} (\bibinfo{year}{2016}).

\bibitem{gardner1998artificial}
\bibinfo{author}{Gardner, M.~W.} \& \bibinfo{author}{Dorling, S.}
\newblock \bibinfo{title}{Artificial neural networks (the multilayer perceptron)—a review of applications in the atmospheric sciences}.
\newblock \emph{\bibinfo{journal}{Atmos Environ}} \textbf{\bibinfo{volume}{32}}, \bibinfo{pages}{2627--2636} (\bibinfo{year}{1998}).

\bibitem{davis2006relationship}
\bibinfo{author}{Davis, J.} \& \bibinfo{author}{Goadrich, M.}
\newblock \bibinfo{title}{The relationship between precision-recall and roc curves}.
\newblock In \emph{\bibinfo{booktitle}{In: Proceedings of the 23rd International Conference on Machine Learning}}, \bibinfo{pages}{233--240} (\bibinfo{year}{2006}).

\bibitem{saito2015precision}
\bibinfo{author}{Saito, T.} \& \bibinfo{author}{Rehmsmeier, M.}
\newblock \bibinfo{title}{The precision-recall plot is more informative than the roc plot when evaluating binary classifiers on imbalanced datasets}.
\newblock \emph{\bibinfo{journal}{PLoS One}} \textbf{\bibinfo{volume}{10}}, \bibinfo{pages}{e0118432} (\bibinfo{year}{2015}).

\bibitem{aral2009distinguishing}
\bibinfo{author}{Aral, S.}, \bibinfo{author}{Muchnik, L.} \& \bibinfo{author}{Sundararajan, A.}
\newblock \bibinfo{title}{Distinguishing influence-based contagion from homophily-driven diffusion in dynamic networks}.
\newblock \emph{\bibinfo{journal}{Proc Natl Acad Sci U S A}} \textbf{\bibinfo{volume}{106}}, \bibinfo{pages}{21544--21549} (\bibinfo{year}{2009}).

\bibitem{lloyd2005superspreading}
\bibinfo{author}{Lloyd-Smith, J.~O.}, \bibinfo{author}{Schreiber, S.~J.}, \bibinfo{author}{Kopp, P.~E.} \& \bibinfo{author}{Getz, W.~M.}
\newblock \bibinfo{title}{Superspreading and the effect of individual variation on disease emergence}.
\newblock \emph{\bibinfo{journal}{Nature}} \textbf{\bibinfo{volume}{438}}, \bibinfo{pages}{355--359} (\bibinfo{year}{2005}).

\bibitem{godri2020covid}
\bibinfo{author}{Godri~Pollitt, K.~J.} \emph{et~al.}
\newblock \bibinfo{title}{Covid-19 vulnerability: the potential impact of genetic susceptibility and airborne transmission}.
\newblock \emph{\bibinfo{journal}{Hum Genomics}} \textbf{\bibinfo{volume}{14}}, \bibinfo{pages}{1--7} (\bibinfo{year}{2020}).

\bibitem{liu2021efficient}
\bibinfo{author}{Liu, Y.} \emph{et~al.}
\newblock \bibinfo{title}{Efficient network immunization under limited knowledge}.
\newblock \emph{\bibinfo{journal}{Natl Sci Rev}} \textbf{\bibinfo{volume}{8}}, \bibinfo{pages}{nwaa229} (\bibinfo{year}{2021}).

\bibitem{tacchella2012new}
\bibinfo{author}{Tacchella, A.}, \bibinfo{author}{Cristelli, M.}, \bibinfo{author}{Caldarelli, G.}, \bibinfo{author}{Gabrielli, A.} \& \bibinfo{author}{Pietronero, L.}
\newblock \bibinfo{title}{A new metrics for countries' fitness and products' complexity}.
\newblock \emph{\bibinfo{journal}{Sci Rep}} \textbf{\bibinfo{volume}{2}}, \bibinfo{pages}{1--7} (\bibinfo{year}{2012}).

\end{thebibliography}

\end{document}